# Investigation of Ta$_2$O$_5$ as an alternative high κ dielectric for InAlN/GaN MOS HEMT on Si

Sandeep kumar, Himanshu Kumar, Sandeep Vura, Anamika Singh Pratiyush, Vanjari Sai Charan, Surani B. Dolmanan, Sudhiranjan Tripathy, Rangarajan Muralidharan, Digbijoy N. Nath *Member* IEEE

*Abstract* - We report on the demonstration and investigation of Ta$_2$O$_5$ as high-κ dielectric for InAlN/GaN-MOS HEMT-on-Si. Ta$_2$O$_5$ of thickness 24 nm and dielectric constant ~ 30 was sputter deposited on InAlN/GaN HEMT and was investigated for different post deposition anneal conditions (PDA). The gate leakage was 16nA/mm at -15 V which was ~ 5 orders of magnitude lower compared to reference HEMT. The 2-dimensional electron gas (2DEG) density was found to vary with annealing temperature suggesting the presence of net charge at the Ta$_2$O$_5$/InAlN interface. Dispersion in the capacitance-voltage (C-V) characteristics was used to estimate the frequency-dependent interface charge while energy band diagrams under flat band conditions were investigated to estimate fixed charge. The optimum anneal condition was found to be 500° C which has resulted into a flat band voltage spread (ΔV$_{FB}$) of 0.4 V and interface fix charge (Q$_f$) of 3.98×10$^{13}$ cm$^{-2}$. XPS (X-ray photoelectron spectroscopy) spectra of as deposited and annealed Ta$_2$O$_5$ film were analyzed for Ta and O compositions in the film. The sample annealed at 500° C has shown Ta:O ratio of 0.41. XRD (X-ray diffraction) analysis was done to check the evolution of poly-crystallization of the Ta$_2$O$_5$ film at higher annealing temperatures.

## I. INTRODUCTION

III-Nitride HEMTs have been investigated for over two decades for their high frequency and high power applications [1]. There is an increasing market penetration of GaN HEMTs for power amplification and power conversion applications including those in mobile base stations, SATCOM/TV, radars, converters/inverters for a wide range of power systems and motor drives. Although AlGaN/GaN HEMTs with 20-30 nm barrier are more widely studied devices including the ones used in the commercial sector, InAlN/GaN and AlN/GaN HEMTs offer several advantages such as very high 2DEG density leading to higher current and ultra-thin barrier leading to higher transconductance (g$_m$) and f$_t$ [2][3][4][5][6]. These thinner barrier devices suffer from higher gate leakage current which necessitates the use of a gate dielectric at the expense of g$_m$. Lattice matched [7] In$_{0.17}$Al$_{0.83}$N/GaN grown on GaN is relatively stress free. High κ dielectric materials such as Al$_2$O$_3$, HfO$_2$ etc were used to reduce the leakage keeping the thickness as low as possible. High power switching applications necessitates very low leakage current and hence use of thicker dielectric with very high-k such as Ta$_2$O$_5$ (ϵ = 20 to 50) [8] can offer a significant advantage. Although Ta$_2$O$_5$ as gate dielectric has been reported AlN/GaN HEMTs [9], Ta$_2$O$_5$/InAlN/GaN HEMT is not yet reported. In this letter, we report on the fabrication and characterization of Ta$_2$O$_5$/InAlN/GaN MOS-HEMT with I$_{ON}$/I$_{OFF}$ ratio >10$^8$ and investigate the Ta$_2$O$_5$/InAlN interface through C-V and band diagram analysis. The material properties of as-deposited and annealed film were investigated using XPS and XRD.

## II. EXPERIMENTAL DETAILS

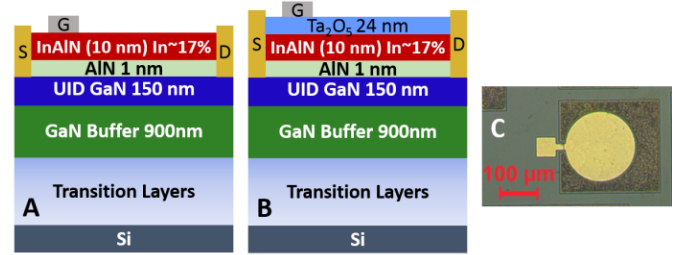

Fig. 1. Epitaxially grown HEMT stack, schematic of (A) HEMT (B) MOS-HEMT (C) optical microscope image of circular Schottky C-V pad.

Figure 1 (A) and Fig. 1 (B) shows the epi-stack with the schematic of HEMT and MOS-HEMT used in this work. The growth details of the epi-stack is published elsewhere [10]. Four samples (three for MOS-HEMT and one for reference HEMT) were co-processed for this study. Device fabrication started with e-beam evaporation of Ti/Al/Ni/Au for source-drain Ohmic contacts followed by lift-off. Rapid thermal anneal was done for 30 s at 850° C in N$_2$ ambient. Samples were mesa etched using Cl$_2$-BCl$_3$ RIE chemistry. Low power O$_2$ plasma for 20 s, NH$_4$OH dip for 5 min and HF-HCl dip for 30 s were used to clean the sample surface before depositing the Ta$_2$O$_5$ oxide. 24 nm of Ta$_2$O$_5$ were sputter deposited on the samples at room temperature. The three samples with Ta$_2$O$_5$ as gate dielectric were subjected to three different post deposition anneal (PDA) conditions of 500° C, 600° C and 700° C for 1 min, in forming gas (FGA). Ni/Au metal was evaporated for gate and all the samples (MOS-HEMT and HEMT) were

S. Kumar, H. Kumar, S. Vura, A. S. Pratiyush, V. S. Charan, R. Muralidharan, and Digbijoy N. Nath are with Centre for Nano Science and Engineering (CeNSE), Indian Institute of Science (IISc), Bengaluru, India (e-mail: sandeepku@iisc.ac.in, digbijoy@iisc.ac.in).

B. Dolmanan and S. Tripathy are with b Institute of Materials Research and Engineering (IMRE), Agency for Science, Technology, and Research (A*STAR), Singapore subjected to a post metal anneal (PMA) in forming gas ambiance at 400° C for 5 min. Source-drain contacts were exposed by selectively etching $Ta_2O_5$ in buffered hydrofluoric (BHF) acid solution. The gate length for transistors was 3 µm while the gate-source and gate-drain spacing were 3 µm and 15 µm respectively. Capacitance voltage (C-V) characteristics were measured on circular Schottky C-V pads (Fig. 1 (C)) of radius 90 µm.

## III. RESULTS AND DISCUSSIONS

DC current voltage (I-V) and C-V measurements were performed to characterize the $Ta_2O_5$ as a dielectric for MOS-HEMT. From transfer length measurements for the sample without oxide, contact resistance ($R_C$) and sheet resistance ($R_{SH}$) were found to be 0.35 Ω-mm and 241 ohm/□ respectively. Gate drain leakage was found to be relatively high ~ 2.6 mA/mm ($V_G$=-15 V) for the control sample (without oxide) (Fig. 2) which is expected of thinner barrier HEMTs. It can be seen from Fig. 2 that the use of $Ta_2O_5$ as gate dielectric led to a reduction of the gate leakage by 5 orders of magnitude (2.6 mA/mm to 16nA/mm, at 15 V) for PDA conditions 500° C and 600° C. High temperature anneal results in poly crystallization of $Ta_2O_5$ film and the main reason behind increased gate leakage compared to 500° C annealed samples. The poly-crystallization of $Ta_2O_5$ at higher annealing temperatures has been confirmed using XRD which is discussed later. The grain boundaries in the polycrystalline film serve as current leakage path [11].

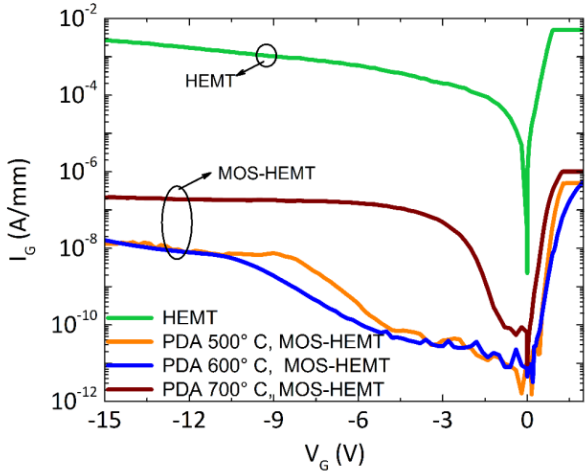

Fig. 2. Gate-Drain leakage current of HEMT and MOS-HEMT (Lgs=3 µm, Lg=3 µm, Lgd=15 µm) with different PDA conditions.

The devices were further analyzed for DC steady state current characteristics. Transfer characteristic ($I_D$-$V_{GS}$) and output characteristics ($I_D$-$V_{DS}$) are shown in the Fig. 3 and Fig. 4 respectively. The zero-bias saturation current (for $V_D$>$Vk_{nee}$ and $V_G$=0 V) was found to be between 500-700 mA/mm while the pinch-off was found to be between -6 to -11 V for all the samples indicating a variation of 2DEG density with PDA conditions. Different annealing conditions results in different interface-states/charges at oxide-nitride interface [12] which leads to modification of net 2DEG of the device. The shift in the pinch-off voltage towards more (less) negative voltage is a result of net decrease (increase) of negative charges at the $Ta_2O_5$/InAlN interface (Fig. 3). Sample with PDA condition of 500° C exhibited the highest On/Off ratio > $10^7$ with a $V_{TH}$ of -9.7 V which is comparable to those reported for MOS-HEMTs elsewhere [13][14]. Due to variation in $V_{TH}$ of the devices, $I_{ON}/I_{OFF}$ ratios were calculated using $I_D$-$V_{DS}$ characteristics (Fig. 4) at gate voltage overdrive ($V_g$-$V_{TH}$) of 6 V and the $I_{ON}/I_{OFF}$ ratios were > $10^7$, $10^5$ and $10^6$ for the samples annealed at 500° C, 600° C, and 700° C respectively.

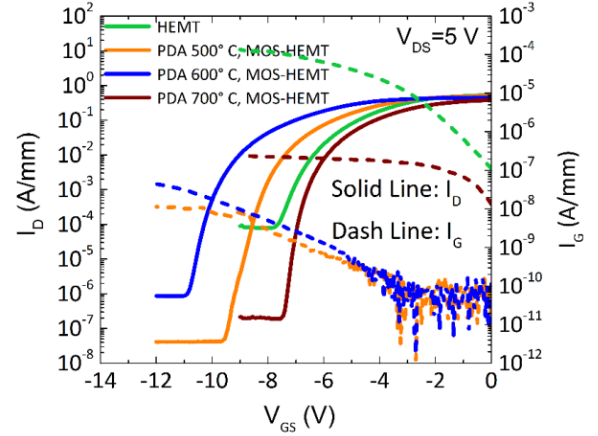

Fig. 3. Drain current and Gate current of HEMT and MOS-HEMT (Lgs=3 µm, Lg=3 µm, Lgd=15 µm) with different PDA conditions.

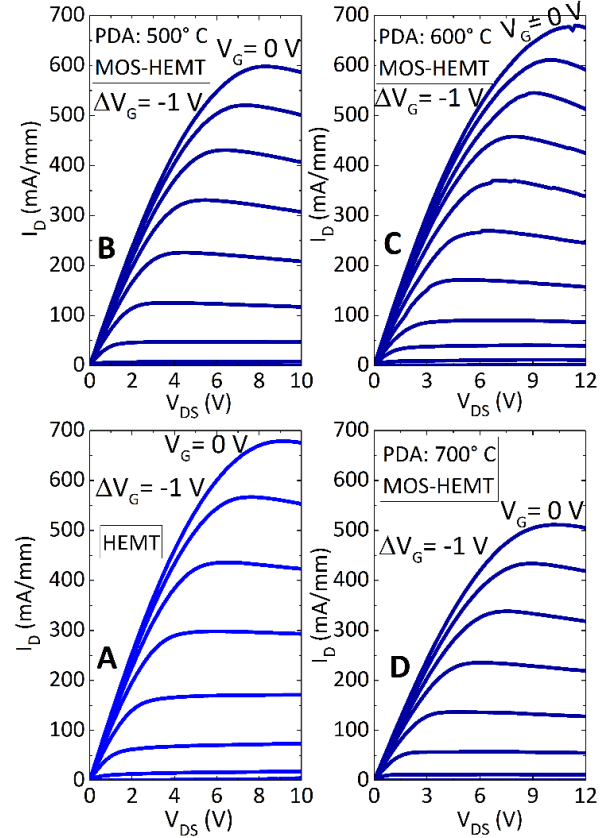

Fig. 4. DC $I_{DS}$-$V_{DS}$ characteristics of HEMT and MOS-HEMT (Lgs=3 µm, Lg=3 µm, Lgd=15 µm) with different PDA conditions. (A) HEMT, (B) MOS-HEMT with PDA 500°C, (C) MOS-HEMT with PDA 600°C, (D) MOS-HEMT with PDA 700°C.

Capacitance voltage (C-V) characteristics of MOS HEMTs and HEMT are shown in Fig. 5 and inset to the Fig. 5 respectively. These measurements were performed on circular Schottky pads. $V_{TH}$ variation in C-V was observed for the samples analogous to that observed for $I_{DS}$-$V_{GS}$. Parameters summarizing the C-V results for both HEMT and MOS-HEMT are listed Table I. The zero-bias capacitance ($C_T$) of MOS-HEMT can be expressed as: $C_T=(C_{Ox}C_{Barr})/(C_{Ox}+C_{Barr})$, where, $C_{Ox}$ and $C_{Barr}$ are capacitances related to $Ta_2O_5$ and InAlN respectively. Using the zero-bias capacitance of HEMT (617 nF/cm$^2$) and the net barrier thickness (11 nm, InAlN+AlN), the relative dielectric constant ($\epsilon$=7.65) of the InAlN barrier was calculated. Then, the $\epsilon$ of $Ta_2O_5$ was estimated using MOS-HEMT and HEMT zero bias capacitances and the estimated values are listed in the Table I. Some uncertainty in the determination of dielectric constant creeps in due to possible errors in the estimation of thicknesses of both InAlN barrier and $Ta_2O_5$ (growth and deposition related un uniformity). However, no significant change (~ ± 0.5 nm) in oxide thickness of the $Ta_2O_5$ on Si (calibration sample) has been observed after post annealing. The calculated threshold voltage ($V_{TH}=qn_s/C_T$, $n_s$ (=2.2×10$^{13}$cm$^{-2}$) is 2DEG density and $C_T$ is total capacitance considering oxide ($Ta_2O_5$=24 nm) and barrier contributions (InAlN=11nm) for MOS HEMT are tabulated in Table 1. This calculated $V_{TH}$ for MOS-HEMT was found to be not in agreement with the experimental $V_{TH}$ (x-axis intercept of the slope line drawn on C-V data) observations. The deviation of experimental $V_{TH}$ from calculated $V_{TH}$ suggests the presence/modulation of charges at $Ta_2O_5$/InAlN interface under different annealing conditions.

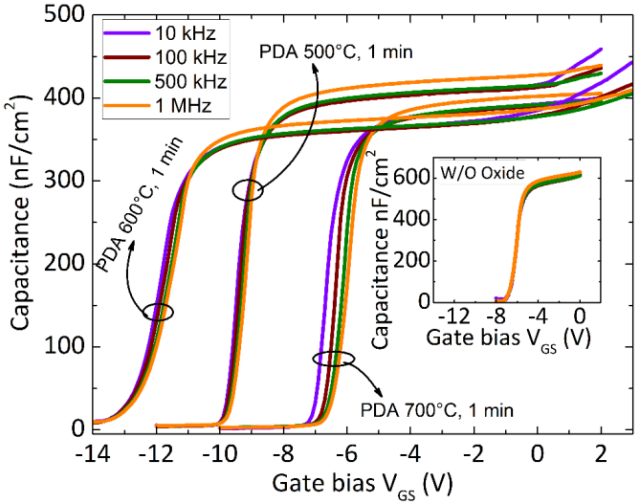

Fig. 5. Capacitance voltage characteristics of HEMT (inset) and MOS-HEMT at different frequencies. Measurements were performed on circular Schottky C-V pads.

Figure 6 shows the schematic band diagram of MOS-HEMT in the pinch-off condition with the polarization charge ($Q_\pi$) and fix-charge ($Q_f$) shown as inset. Here, the fixed charge at the $Ta_2O_5$/InAlN interface was estimated using the polarization charge, band offset values and Schottky barrier height of metal oxide junction. The spontaneous polarization charge[15] in $In_{0.17}Al_{0.83}N$ and GaN are 4.54×10$^{13}$ cm$^{-2}$ and 1.81×10$^{13}$ cm$^{-2}$ respectively. Conduction band offset at $In_{0.17}Al_{0.83}N$/$Ta_2O_5$ was calculated by electron affinity rule considering the electron affinity of $In_{0.17}Al_{0.83}N$ and $Ta_2O_5$ as 2.68 [16] and 3.2 [17] respectively. The conduction band offset considered at InAlN/GaN interface was 0.46 eV [18]. Schottky barrier height ($\varphi_B$) of 0.72 eV [19] was used for Ni/$Ta_2O_5$ interface and for UID GaN (1×10$^{16}$) the estimated value of $E_C$-$E_F$ was 0.14 eV. The expression of flat band voltage for Fig. 6 can be written by the following equation (1):

$$-qV_{FB} + \phi_B = qV_{Ox} + qV_{InAlN} \\ -\Delta Ec_{ox/InAlN} + \Delta Ec_{InAlN/GaN} \\ +(E_C - E_F) \quad (1)$$

where,

$$V_{ox} = q(Q_f - Q_{\pi\_GaN}) \times \frac{T_{ox}}{\varepsilon_0 \varepsilon_{ox}} \quad (2)$$

$$V_{InAlN} = q(Q_{\pi\_InAlN} - Q_{\pi\_GaN}) \times \frac{T_{InAlN}}{\varepsilon_0 \varepsilon_{InAlN}} \quad (3)$$

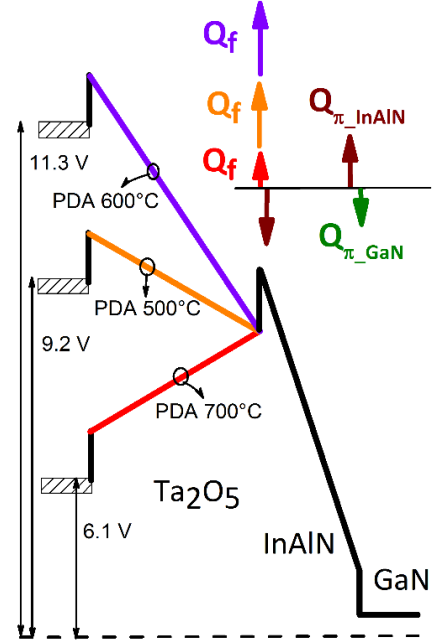

Fig. 6. Capacitance voltage characteristics of HEMT (inset) and MOS-HEMT at different frequencies. Measurements were performed on circular Schottky C-V pads.

Using the above equations (1-3), the fixed charges ($Q_f$) at the InAlN/$Ta_2O_5$ interface for the three samples were estimated as shown in Table I. The fixed charge estimated for the sample (PDA 500° C) was 3.98×10$^{13}$ cm$^{-2}$ which compares well with that reported elsewhere (1-6×10$^{13}$ cm$^{-2}$) [20][21][22] for $Al_2O_3$/AlN, $Al_2O_3$/GaN and SiNx/InAlN. The increase in 2DEG density with increase in the positive interface charge has also been reported previously [21]. The PDA condition dependent fixed charges at $Ta_2O_5$/InAlN interface were calculated by considering constant conduction band offset at this interface. However, the dipole formation at the interfaces in the presence of fixed charges can change the band offsets [23]. Different characterization techniques would need to be investigated to estimate the band offset variations with charges.

Two step C-V characteristics have been widely used to characterize dielectric/barrier and barrier/channel interfaces of MOS HEMTs. Using the conductance method, the $G/\omega$ versus frequency characteristics near $V_{TH}$ and near second slope in C-V (for $V_G>0$ V) were used to estimate the trap densities at barrier/channel and dielectric/barrier interfaces respectively [24][25][26]. The band alignment between widely reported barrier ($Al_{0.25}Ga_{0.75}N$) and dielectrics ($Al_2O_3$, $SiO_2$, $HfO_2$) supports the electron blocking at dielectric/barrier interface. In forward bias condition ($V_G>0$ V), the electrons get transferred from barrier/channel interface to the dielectric/barrier interface. These electrons neutralize the donor traps and make the acceptor type traps negatively charged as a result the second slope in C-V is observed for $V_G > 0$ V [24]. Here in this case, the band alignment (Fig. 6) of $Ta_2O_5$/InAlN doesn't support electron blocking for positive gate bias due to this reason the conventional conductance method was not employed for trap characterization. As the conduction band offset of $Ta_2O_5$/InAlN does not support electron blocking at this interface, $Al_2O_3$/$Ta_2O_5$ stack on InAlN may be employed to block electrons and the $Ta_2O_5$/InAlN interface can be characterized using the conventional conductance method. Due to these restrictions, C-V measurements (Fig. 5) at various frequencies and the spread in C-V characteristics should be employed for characterizing frequency dependent properties of dielectric and interfaces. The HEMT C-V characteristics show very less spread (difference of flat band voltages ($\Delta V_{FB}$) at 10 kHz and 1 MHz was 0.15 V). This negligible spread in C-V characteristics (Fig.5) is an indication of good interface between InAlN barrier and GaN channel. The relatively higher spread in frequency-dependent capacitances of MOS-HEMTs could be attributed to the interface state present at $Ta_2O_5$/InAlN interface or in $Ta_2O_5$. A measure of spread in C-V characteristics can be defined as the difference in flat band voltages at 10 kHz and 1 MHz. This flat band voltage difference ($\Delta V_{FB}$) for MOS HEMT and HEMT is shown in Table I. $\Delta V_{FB}$ varies from 0.15 V to 0.8 V for HEMT and MOS HEMT (PDA 700° C) respectively. The least value of $\Delta V_{FB}$ observed for MOS HEMT (PDA 500° C) was 0.4 V.

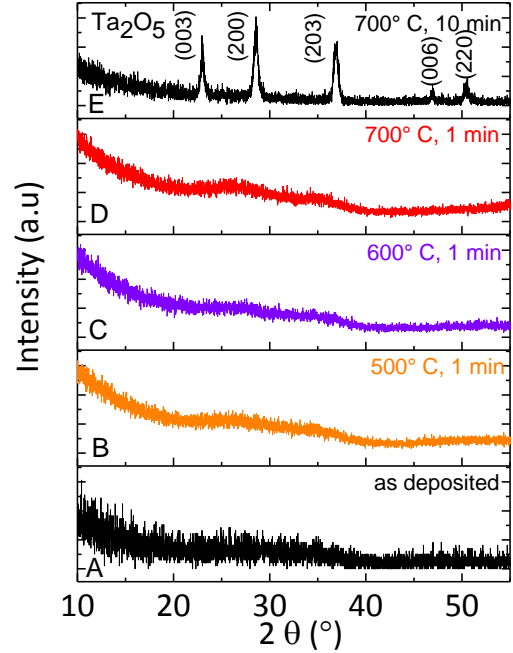

Fig. 7. XRD 2θ scans of $Ta_2O_5$ films of thickness 55 nm (A) as-deposited, FGA at (B) 500° C for 1 min (C) 600° C for 1 min (D) 700° C for 1 min (E) 700° C for 10 min.

Figure 7 shows the XRD 2θ scan of the $Ta_2O_5$ films (as deposited and annealed under different conditions) on HEMT. The absence of any Bragg reflection in the as-deposited film confirms its amorphous nature. When $Ta_2O_5$ films were subjected to rapid thermal annealing in forming gas at 500° C, 600° C, 700° C for 1 min, 2θ scans started to show humps at 2θ positions of 28° and 35° (Fig. 7 (B, C, D)). Similar results have been observed in extended X-ray absorption florescence spectroscopic (EXAFS) studies [27]. In-order to confirm that $Ta_2O_5$ films were in process of crystallization, one sample was annealed at 700° C for 10 mins (the 2θ scan is shown Fig. 7(E)). The Bragg reflections from the planes can be indexed as (003), (200), (203), (006), (220) of α-$Ta_2O_5$ (hexagonal) (PDF-00-018-1304). The poly crystallization of $Ta_2O_5$ films was confirmed by XRD for the samples annealed at higher temperatures (~700° C) and it is in agreement with previous report [28].

TABLE I
C-V CHARACTERISTICS FOR DIFFERENT PDA CONDITIONS

| PDA Conditions | $V_{FB}$ FLAT BAND VOLTAGE (V) @ 500 KHZ | 2DEG calculated from C-V (@ 500 kHz) (cm$^{-2}$) | ε of $Ta_2O_5$ (calculated) | $V_{TH}$(V) calculated using ε of $Ta_2O_5$ and $n_s$=2.2x10$^{13}$ cm$^{-2}$ | $V_{TH}$ (V) experimental @500 kHz | $\Delta V_{FB}$ (10 kHz to 1MHz) | $Q_f$ (cm$^{-2}$) for Delta $E_C$=0.52 eV |
|---|---|---|---|---|---|---|---|
| MOS-HEMT, 500° C | -9.2 | 2.3×10$^{13}$ | 34.5 | 8.5 | -9.7 | 0.4 V | 3.98×10$^{13}$ |
| MOS-HEMT, 600° C | -11.3 | 2.64×10$^{13}$ | 26.3 | 9.3 | -12.6 | 0.6 V | 4.74×10$^{13}$ |
| MOS HEMT, 700° C | -6.1 | 1.45×10$^{13}$ | 29.2 | 8.9 | -6.6 | 0.8 V | 1.56×10$^{13}$ |
| HEMT | -5.95 | 2.2×10$^{13}$ | -- | -- | -6.8 | 0.15 V | -- |

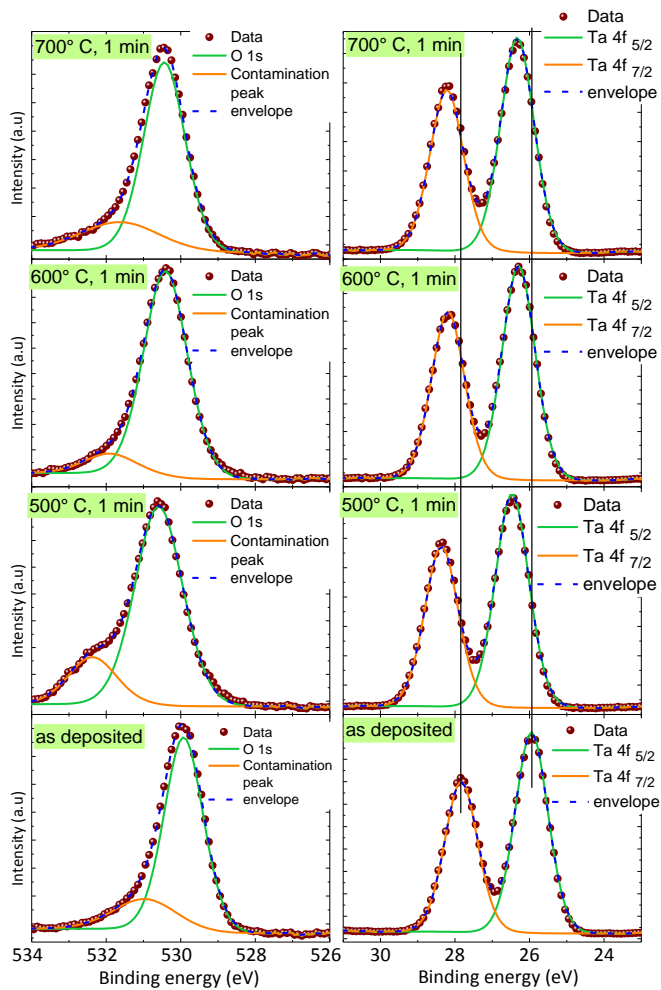

Fig. 8. O (1s) and Ta (4f) core level XPS spectrum of the samples (9 nm thick $Ta_2O_5$ on InAlN/GaN) with different PDA conditions

Figure 8 shows the XPS spectrum of O (1s) and Ta (4f) for different anneal conditions. The XPS spectra was calibrated using C 1s peak position (284.6 eV). The thickness of the $Ta_2O_5$ film used for XPS was 9 nm. Ta-O binding energy of $Ta_2O_5$ is characterized by the doublets (peaks corresponding to $4f_{5/2}$ and $4f_{7/2}$) observed in the Ta 4f spectra and these peaks have a separation of 1.89 eV [29]. The peak of XPS spectra shifts by ~0.47 eV (towards higher binding energy) for 500° C annealed sample compared to as-deposited sample. The peak shifts were 0.35 eV and 0.40 eV for 600° C and 700 °C annealed samples respectively. The shift of the peaks with different anneal conditions were used to quantify the $Ta_2O_5$, a fully oxidized film corresponds to higher binding energy peak and a partial oxidation corresponds to lower binding energy [30]. The $Ta_2O_5$ film was further analyzed for atomic compositions of Ta and O by analyzing the Ta and O spectra. The ratio of Ta:O in $Ta_2O_5$ is 0.4 (2/5) for the ideal film and this ratio was found to be 0.64 for as deposited film. After the FGA, the ratio has changed to 0.41, 0.43 and 0.52 for 500° C, 600° C and 700° C annealed samples respectively. 500° C annealing condition has resulted in better stochiometric film (close to ideal Ta:O ratio) while 600° C and 700° C annealed films have become oxygen deficient (Ta rich). XPS spectra of Ta was dominated with $Ta^{5+}$ state in all the samples [31][32].

The O (1s) spectra were mostly clean for all the samples with surface contamination signatures (~532 eV) [31].

## IV. CONCLUSIONS

We have demonstrated $Ta_2O_5$/InAlN/GaN MOS HEMT on Si. The dielectric constant of $Ta_2O_5$ was found to be ~30. Different anneal conditions were found to affect the net 2DEG density. Under different PDA conditions, fix charge at $Ta_2O_5$/InAlN interface and $\Delta V_{FB}$ were estimated. The measure of spread in C-V, $\Delta V_{FB}$ was found least for 500° C PDA sample. XPS spectra of the $Ta_2O_5$ film suggest the improvement in oxide quality for 500° C annealing and with higher annealing temperatures (600° C and 700° C) the oxide quality deteriorates. Annealing at higher temperature leads to poly crystallization of $Ta_2O_5$ as a result gate leakage current increases. The results presented here show the promise of using $Ta_2O_5$ as a very high κ dielectric with dielectric/nitride interface of comparable quality to most of the reported dielectric/nitride interfaces. However, quantitative trap densities estimation and XPS analysis of $Ta_2O_5$/InAlN (Nitride) interface needs to be done to establish $Ta_2O_5$ as an alternative high κ dielectric for nitride HEMTs.

This publication is an outcome of the Research and Development work undertaken in the Project under Ph.D. scheme of Media Lab Asia. We acknowledge funding support from MHRD through NIEIN project, from MeitY and DST through NNetRA. Authors would also like to acknowledge the National NanoFabrication Centre (NNFC) and Micro and Nano Characterization Facility (MNCF) at CeNSE, IISc for device fabrication and characterization.

## REFERENCES

[1] U. K. Mishra, P. Parikh, and Y. F. Wu, "AlGaN/GaN HEMTs - An overview of device operation and applications," *Proc. IEEE*, vol. 90, no. 6, pp. 1022–1031, 2002.

[2] J. Kuzmík, "Power electronics on InAlN/(In)GaN: Prospect for a record performance," *IEEE Electron Device Lett.*, vol. 22, no. 11, pp. 510–512, 2001.

[3] F. Medjdoub, J. F. Carlin, M. Gonschorek, E. Feltin, M. A. Py, D. Ducatteau, C. Gaquière, N. Grandjean, and E. Kohn, "Can InAlN/GaN be an alternative to high power / high temperature AlGaN/GaN devices?," *Tech. Dig. - Int. Electron Devices Meet. IEDM*, no. 1, pp. 1–4, 2006.

[4] D. Deen, T. Zimmermann, Y. Cao, D. Jena, and H. G. Xing, "2.3 nm barrier AlN/GaN HEMTs with insulated gates," *Phys. Status Solidi Curr. Top. Solid State Phys.*, vol. 5, no. 6, pp. 2047–2049, 2008.

[5] Y. Cao, K. Wang, and D. Jena, "Electron transport properties of low sheet-resistance two-dimensional electron gases in ultrathin AlN/GaN heterojunctions grown by MBE," *Phys. Status Solidi Curr. Top. Solid State Phys.*, vol. 5, no. 6, pp. 1873–1875, 2008.

[6] R. S. Pengelly, *Microwave Field-Effect Transistors Theory, Design, and Applications.* Research studies press, Letchworth, England, 1986.

[7] S. Schmult, T. Siegrist, A. M. Sergent, M. J. Manfra, and R. J. Molnar, "Optimized growth of lattice-matched InxAl1-xN/GaN heterostructures by molecular beam epitaxy," *Appl. Phys. Lett.*, vol. 90, p. 21922, 2007.

[8] C. Chaneliere, J. L. Autran, R. a. B. Devine, and B. Balland, "Tantalum pentoxide (Ta2O5) thin films for advanced dielectric applications," *Mater. Sci. Eng. R Reports*, vol. 22, no. 6, pp. 269–322, 1998.

[9] D. A. Deen, D. F. Storm, R. Bass, D. J. Meyer, D. S. Katzer, S. C. Binari, J. W. Lacis, and T. Gougousi, "Atomic layer deposited

Ta2O5 gate insulation for enhancing breakdown voltage of AlN/GaN high electron mobility transistors," *Appl. Phys. Lett.*, vol. 98, no. 2, pp. 2009–2012, 2011.

[10] S. Kumar, A. S. Pratiyush, S. B. Dolmanan, and S. Tripathy, "UV Detector based on InAlN/GaN-on-Si HEMT Stack with photo-to-dark current ratio > 107," *Appl. Phys. Lett.*, vol. 111, p. 251103, 2017.

[11] T. Kubo, J. J. Freedsman, Y. Iwata, and T. Egawa, "Electrical properties of GaN-based metal-insulator-semiconductor structures with Al2O3 deposited by atomic layer deposition using water and ozone as the oxygen precursors," *Semicond. Sci. Technol.*, vol. 29, p. 45004, 2014.

[12] H. Zhou, G. I. Ng, Z. H. Liu, and S. Arulkumaran, "Improved device performance by post-oxide annealing in atomic-layer-deposited Al2O3/AlGaN/GaN metal-insulator-semiconductor high electron mobility transistor on Si," *Appl. Phys. Express*, vol. 4, no. 10, pp. 3–6, 2011.

[13] B. Y. Chou, H. Y. Liu, W. C. Hsu, C. S. Lee, Y. S. Wu, and E. P. Yao, "Electrical and reliability performances of stacked HfO2/Al2O3 MOS-HEMTs," *Int. Conf. Inf. Sci. Electron. Electr. Eng. Sapporo*, pp. 977–981, 2014.

[14] A. Chakroun, A. Jaouad, A. Soltani, O. Arenas, V. Aimez, R. Arès, and H. Maher, "AlGaN/GaN MOS-HEMT device fabricated using a high quality PECVD passivation process," *IEEE Electron Device Lett.*, vol. 38, no. 6, pp. 779–782, 2017.

[15] F. Bernardini, V. Fiorentini, and D. Vanderbilt, "Spontaneous polarization and piezoelectric constants of III-V nitrides," *Phys. Rev. B*, vol. 56, no. 16, pp. 24–27, 1997.

[16] J. Kuzmík, "InAlN/(In)GaN high electron mobility transistors: some aspects of the quantum well heterostructure proposal," *Semicond. Sci. Technol.*, vol. 17, pp. 540–544, 2002.

[17] B. Chih-ming Lai, N. Kung, and J. Ya-min Lee, "A study on the capacitance–voltage characteristics of metal-Ta2O5-silicon capacitors for very large scale integration metal-oxide-semiconductor gate oxide applications," *J. Appl. Phys.*, vol. 85, no. 8, pp. 4087–4090, 1999.

[18] M Grundmann, "'BandEngineering.'http://my.ece.ucsb.edu/mgrundmann/bandeng.htm." .

[19] J. Robertson and B. Falabretti, "Band offsets of high K gate oxides on III-V semiconductors," *J. Appl. Phys.*, vol. 100, p. 14111, 2006.

[20] G. Dutta, S. Turuvekere, N. Karumuri, N. Dasgupta, and A. Dasgupta, "Positive shift in threshold voltage for reactive-ion-sputtered Al2O3/AlInN/GaN MIS-HEMT," *IEEE Electron Device Lett.*, vol. 35, no. 11, pp. 1085–1087, 2014.

[21] S. Ganguly, J. Verma, G. Li, T. Zimmermann, H. Xing, and D. Jena, "Presence and origin of interface charges at atomic-layer deposited Al2O3/III-nitride heterojunctions," *Appl. Phys. Lett.*, vol. 99, no. 19, pp. 2009–2012, 2011.

[22] M. Esposto, S. Krishnamoorthy, D. N. Nath, S. Bajaj, T. H. Hung, and S. Rajan, "Electrical properties of atomic layer deposited aluminum oxide on gallium nitride," *Appl. Phys. Lett.*, vol. 99, no. 13, pp. 2–5, 2011.

[23] J. Robertson, "Band offsets of high dielectric constant gate oxides on silicon," *J. Non. Cryst. Solids*, vol. 303, no. 1, pp. 94–100, 2002.

[24] C. Mizue, Y. Hori, M. Miczek, and T. Hashizume, "Capacitance–Voltage Characteristics of Al$_2$O$_3$/AlGaN/GaN Structures and State Density Distribution at Al$_2$O$_3$/AlGaN Interface," *Jpn. J. Appl. Phys.*, vol. 50, no. 2, p. 21001, 2011.

[25] S. Kumar, N. Remesh, S. B. Dolmanan, S. Tripathy, S. Raghavan, R. Muralidharan, and D. N. Nath, "Interface traps at Al$_2$O$_3$/InAlN/GaN MOS-HEMT-on-200 mm Si," *Solid State Electron.*, vol. 137, pp. 117–122, 2017.

[26] S. Kumar, P. Gupta, I. Guiney, C. J. Humphreys, S. Raghavan, R. Muralidharan, and D. N. Nath, "Temperature and Bias Dependent Trap Capture Cross Section in AlGaN/GaN HEMT on 6-in Silicon With Carbon-Doped Buffer," *IEEE Trans. Electron Devices*, 2017.

[27] R. V. Gonçalves, P. Migowski, H. Wender, A. F. Feil, M. J. M. Zapata, S. Khan, F. Bernardi, G. M. Azevedo, and S. R. Teixeira, "On the crystallization of Ta$_2$O$_5$ nanotubes: structural and local atomic properties investigated by EXAFS and XRD," *CrystEngComm*, vol. 16, pp. 797–804, 2014.

[28] T. Yoshimoto, T. Goto, H. Takagi, Y. Nakamura, H. Uchida, C. A. Ross, and M. Inoue, "Thermally stable amorphous tantalum yttrium oxide with low IR absorption for magnetophotonic devices," *Sci. Rep.*, vol. 7, p. 13805, 2017.

[29] R. S. Devan, C.-L. Lin, S.-Y. Gao, C.-L. Cheng, Y. Liou, and Y.-R. Ma, "Enhancement of green-light photoluminescence of Ta2O5 nanoblock stacks," *Phys. Chem. Chem. Phys.*, vol. 13, p. 13441, 2011.

[30] C.-H. Kao, H. Chen, and C.-Y. Chen, "Material and electrical characterizations of high-k Ta2O5 dielectric material deposited on polycrystalline silicon and single crystalline substrate," *Microelectron. Eng.*, vol. 138, pp. 36–41, 2015.

[31] E. Atanassova, G. Tyuliev, A. Paskaleva, D. Spassov, and K. Kostov, "XPS study of N2 annealing effect on thermal Ta2O5 layers on Si," *Appl. Surf. Sci.*, vol. 225, pp. 86–99, 2004.

[32] R. V. Gonçalves, R. Wojcieszak, P. M. Uberman, S. R. Teixeira, and L. M. Rossi, "Insights into the active surface species formed on Ta2O5 nanotubes in the catalytic oxidation of CO," *Phys. Chem. Chem. Phys.*, vol. 16, p. 5755, 2014.